\documentclass[prd,twocolumn,superscriptaddress,floatfix,nofootinbib]{revtex4-2}
\pdfoutput=1 
\usepackage[T1]{fontenc} 
\usepackage{amsmath,amssymb,amsfonts,amsthm}
\usepackage{graphicx}
\usepackage[usenames,dvipsnames]{color}
\usepackage{enumitem}
\usepackage{verbatim}
\usepackage[percent]{overpic}
\usepackage{rotating}
\usepackage{hyperref}
\usepackage{array}
\usepackage{mathtools}
\usepackage{lmodern}
\usepackage[normalem]{ulem}
\usepackage{braket}
\usepackage{tensor}
\usepackage{dsfont}
\usepackage{bm}
\usepackage{tikz}

\usepackage{mathrsfs}

\usetikzlibrary{decorations.pathmorphing}
\usepackage{CJKutf8}

\usetikzlibrary{positioning}
\usetikzlibrary{calc,through,backgrounds}

\theoremstyle{definition}
\newtheorem{definition}{Definition}[section]
\newtheorem{theorem}{Theorem}[section]

\newcommand{\kako}[1]{\left( #1 \right)}

\newcommand{\ts}[1]{ _{\text{#1}} }

\newcommand{\biggkako}[1]{\bigg( #1 \bigg)}

\allowdisplaybreaks

\DeclareMathOperator{\Tr}{Tr}
\newcommand{\R}{\mathbb{R}}
\newcommand{\dd}{\text{d}}

\newcommand{\bx}{{\bm{x}}}

\newcommand{\id}{\mathds{1}}
\newcommand{\sx}{\mathsf{x}}

\newcommand{\ii}{\mathsf{i}}

\newcommand{\AAA}{\text{A}}
\newcommand{\BB}{\text{B}}

\begin{document}

\title{Correlation harvesting between particle detectors in uniform motion}


\author{Lana Bozanic}
\email{lbozanic@uwaterloo.ca}
\affiliation{Department of Applied Mathematics, University of Waterloo, Waterloo, Ontario, N2L 3G1, Canada}

\author{Manar Naeem}
\email{manar.naeem@uwaterloo.ca}
\affiliation{Department of Physics and Astronomy, University of Waterloo, Waterloo, Ontario, N2L 3G1, Canada}
\affiliation{Institute for Quantum Computing, University of Waterloo, Waterloo, Ontario, N2L 3G1, Canada}

\author{Kensuke Gallock-Yoshimura}
\email{kgallock@uwaterloo.ca} 

\affiliation{Department of Physics and Astronomy, University of Waterloo, Waterloo, Ontario, N2L 3G1, Canada}


\author{Robert B. Mann}
\email{rbmann@uwaterloo.ca}
\affiliation{Department of Physics and Astronomy, University of Waterloo, Waterloo, Ontario, N2L 3G1, Canada}

\begin{abstract}
We investigate the correlation harvesting protocol using two Unruh-DeWitt particle detectors moving along four classes of uniformly accelerated trajectories categorized by Letaw: linear, catenary, cusped, and circular motions. 
For each trajectory, two types of configurations are carried out: one possesses a stationary (time-translation invariant) Wightman function and the other is nonstationary. 
We find that detectors undergoing linear, catenary, and cusped motions gain fewer correlations in the nonstationary configurations compared to those in stationary configurations. 
Detectors in circular motion have similar behavior in both configurations.
We discuss the relative suppression  of correlation harvesting due to high acceleration for each case. 
Remarkably we find that under certain circumstances detectors in both linear and circular states of motion can harvest genuine (non-communication assisted) 
entanglement even though they are in causal contact.

\end{abstract}

\maketitle
\flushbottom

\section{Introduction}
The field of relativistic quantum information (RQI) has undergone rapid development in recent years, resulting in the emergence of groundbreaking concepts such as entanglement degradation due to non-inertial motion \cite{FuentesAliceFalls,AlsingDiracFields,Alsing2012review},
entanglement harvesting \cite{Valentini1991nonlocalcorr, reznik2003entanglement, reznik2005violating, pozas2015harvesting,salton2015acceleration}, and the long-established Unruh effect \cite{PhysRevD.7.2850, davies_1975, Unruh1979evaporation}. 
The Unruh effect states that a linearly accelerating observer in Minkowski spacetime will experience a thermal bath, and this experience is indistinguishable from that of an inertial observer sitting in a thermal bath. 
More precisely, the temperature detected by a linearly accelerating two-level quantum system, known as the Unruh-DeWitt (UDW) particle detector \cite{Unruh1979evaporation, DeWitt1979}, is proportional to its acceleration $a$ and it reads
\begin{equation}
    T\ts{U}= \dfrac{\hbar a}{2\pi c k\ts{B}}\,,
    \label{eqn:UnruhTempeatureLinear}
\end{equation}
where $\hbar$ is the reduced Planck constant, $c$ is the speed of light, and $k\ts{B}$ is  Boltzmann's constant.

Despite its theoretical significance, the Unruh effect has yet to be verified experimentally. 
The main challenge hindering its experimental verification lies in the large acceleration required to produce experimentally measurable temperatures. 
For example, an acceleration on the order of magnitude of $a \approx 10^{20}$ m/s$^2$ is needed to achieve a temperature of $T\ts{U}\sim 1$ Kelvin.

Given this situation,  researchers have explored other detector trajectories that could induce a phenomenon similar to the Unruh effect. 
In 1981, Letaw found five classes of stationary trajectories with nonzero constant acceleration in flat spacetime \cite{Letaw.uniform.acceleration}: linear, circular, cusped, catenary, and helix trajectories. 
In $(3+1)$-dimensional Minkowski spacetime, these trajectories are characterized by three parameters: two torsions and the magnitude of the proper acceleration. 
The effective temperatures observed by a detector undergoing these motions have been studied over subsequent years in various contexts \cite{Bell.circular.1983, Bell.Leinaas.1987, Unruh.acceleration.rad.electron.1998, Juarez.Asymptotic.states.2019, Good.Unruhlike.temperature2020, CircularTemperaturesUnruh, bunney2023circular}. 
Among these, the circular trajectory has attracted considerable attention for potential experimental realizations of the Unruh effect \cite{Bell.circular.1983, Bell.Leinaas.1987, Retzker.BEC.acceleration.rad, Marino.circular.zero-point, BEC_Unruh, CircularTemperaturesUnruh, bunney2023sound}.

 Less studied is the \textit{entanglement harvesting} protocol for these various classes of motion, apart from detectors undergoing linearly accelerated motion, which has been extensively analyzed \cite{salton2015acceleration, Liu.harvesting.reflecting.boundary, Liu:2021dnl, Liu.acceleration.vs.thermal, ManarKenMutual}. 
In this paper, we consider this problem and investigate how   \textit{two} UDW detectors can
extract entanglement from the vacuum whilst undergoing these various types of non-inertial motion.

In the entanglement harvesting protocol \cite{Valentini1991nonlocalcorr, reznik2003entanglement, reznik2005violating, pozas2015harvesting}, two initially uncorrelated detectors interact locally with a quantum field
in some state (typically the vacuum state)
to extract preexisting entanglement \cite{summers1985bell, summers1987bell}. 
More generally, detectors can harvest classical and quantum correlations in what is called the \textit{correlation harvesting protocol}. 
The amount of harvested correlations is sensitive to the background spacetime \cite{Steeg2009, Cliche.harvesting.weakgrav, smith2016topology, kukita2017harvesting, henderson2018harvesting, ng2018AdS, henderson2019entangling, cong2020horizon, robbins2020entanglement, Xu.Grav.waves.PRD.102.065019, Tjoa2020vaidya, Ken.Freefall.PhysRevD.104.025001, FinnShockwave, Kendra.BTZ, Henderson:2022oyd, caribe2023lensing} and the motion of the detectors \cite{Doukas.orbit.PhysRevA.81.062320, salton2015acceleration, Zhang.harvesting.circular, Liu.harvesting.reflecting.boundary, Liu:2021dnl, FooSuperpositionTrajectory, Diki.inertial, Liu.acceleration.vs.thermal, ManarKenMutual}. Implementing this protocol is very close to implementation as recent experiments detecting correlations of the electromagnetic ground state in a ZnTe crystal have demonstrated \cite{Benea.Electric.correlation.experiment, Settembrini.Detection.correlation.experiment.2022, Lindel2023separately.experiment}.

In this paper we investigate the correlation harvesting protocol with detectors in four classes of uniform acceleration motion: linear, catenary, cusped, and circular trajectories. 
Unlike the helical case, these motions can all be realized in 2 spatial dimensions, and so are more amenable to experimental testing \cite{ Retzker.BEC.acceleration.rad, Marino.circular.zero-point, BEC_Unruh, bunney2023circular,CircularTemperaturesUnruh, bunney2023sound}.
We categorize the configurations for the detectors into two configurations: stationary, in which the Wightman function is time-translation invariant, and nonstationary, in which the Wightman function is not time-translation invariant. 
The Wightman functions for these two scenarios are similar, except in nonstationary configurations they possess an additional term that breaks 
time-translation invariance.

After introducing the UDW detector model in section \ref{sec:setup} and the four uniform acceleration trajectories in section \ref{sec: uniform acceleration trajectories}, we then focus in section \ref{subsec:transition prob acceleration} on a single detector following the four trajectories to examine its transition probability (or response function).  We study its dependence on the magnitudes of the acceleration and torsions, and 
 numerically evaluate the  effective temperature of the detector.
 
We then consider the correlation harvesting protocol in section \ref{subsec:Conc and MI vs a and b}. 
Specifically, concurrence of entanglement and quantum mutual information -- which measures the harvested total correlations -- are numerically evaluated. 
We find that the stationary and nonstationary configurations behave in a similar manner since their Wightman functions have terms in common. 
However, the amount of correlations extracted by the detectors in the nonstationary configurations differs from those of the stationary ones due to an additional term in the Wightman function. 
We also look into the acceleration dependence of the harvested correlations and conclude that sufficiently high accelerations prevent \textit{any} uniformly accelerating detectors from extracting correlations. 
This point is consistent with previous papers that focused on linear and circular motions \cite{Doukas.orbit.PhysRevA.81.062320, salton2015acceleration, Zhang.harvesting.circular, Liu.harvesting.reflecting.boundary, Liu:2021dnl, Liu.acceleration.vs.thermal, ManarKenMutual}. 
Finally, we show that constant acceleration makes it challenging to extract `genuine entanglement' (entanglement preexistent in a quantum field that has no possible assistance from detector communication) in section \ref{subsec:genuine entanglement}. 
In general, genuine entanglement can be harvested from causally disconnected spacetime regions due to microcausality. 
For inertial detectors with Gaussian switching in Minkowski spacetime, it is shown that a sufficiently large energy gap allows the detectors to extract genuine entanglement from such regions \cite{pozas2015harvesting, TjoaSignal}. 
While we find that this is generally not the case for uniformly accelerated detectors,
remarkably we find small but non-negligible regions of parameter space where detectors in causal contact can harvest genuine entanglement.

Throughout this manuscript, we use the mostly-plus metric convention, $(-,+,+,+)$ and the natural units $\hbar =k\ts{B}=c=1$. 
A point in spacetime is denoted by $\sx$.

\section{Unruh-DeWitt detectors}
\label{sec:setup}

\subsection{Density matrix of detectors}
\label{subsec:density matrix}

Let us first review the correlation harvesting protocol. 
Consider two pointlike UDW detectors A and B with an energy gap $\Omega_j,\,j\in \{ \AAA, \BB \}$ between ground $\ket{g_j}$ and excited states $\ket{e_j}$. 
These detectors interact with the quantum Klein-Gordon field $\hat\phi$ along their trajectories $\sx_j(\tau_j)=(t(\tau_j), \bx(\tau_j))$, where $\tau_j$ is the proper time of detector-$j$.

In the interaction picture, the interaction Hamiltonian (as a generator of time-translation with respect to $\tau_j$) describing the coupling between detector-$j$ and $\hat \phi$ is given by 
\begin{align}
    \hat H_j^{ \tau_j } ( \tau_j )
    &=
        \lambda_j \chi_j(\tau_j) \hat \mu_j(\tau_j) 
        \otimes \hat \phi(\sx_j(\tau_j))\,,~j\in \{ \AAA, \BB \}
\end{align}
where $\lambda_j$ is a coupling constant and $\chi_j(\tau_j)$ is the switching function that governs the time-dependence of the coupling. 
Here, $\hat \mu_j(\tau_j) $ is the monopole moment given by
\begin{align}
    \hat \mu_j(\tau_j) 
    &=
        \ket{e_j} \bra{g_j} e^{ \ii \Omega_j \tau_j }
        +
        \ket{g_j} \bra{e_j} e^{ -\ii \Omega_j \tau_j }\,,
\end{align}
which describes each detector's internal dynamics, and the field operator $\hat \phi(\sx_j(\tau_j))$ is pulled back along the trajectory of detector-$j$. 
The superscript on $\hat H_j^{ \tau_j } ( \tau_j )$ indicates the time-translation that the Hamiltonian is generating.

The total interaction Hamiltonian, $\hat H\ts{I}^t(t)$, can be written as a generator of time-translation with respect to the time $t$ that is common to both detectors: 
\begin{align}
    \hat H\ts{I}^t(t)
    &=
        \dfrac{\dd \tau\ts{A}}{\dd t} 
        \hat H\ts{A}^{ \tau\ts{A} }\big( \tau\ts{A}(t) \big)
        +
        \dfrac{\dd \tau\ts{B}}{\dd t} 
        \hat H\ts{B}^{ \tau\ts{B} }\big( \tau\ts{B}(t) \big) \,,
\end{align}
Note that the proper times $\tau\ts{A}$ and $\tau\ts{B}$ are each now  functions of $t$. 
From this Hamiltonian, one obtains the time-evolution operator $\hat U\ts{I}$   \cite{EMM.Relativistic.quantum.optics,Tales2020GRQO}: 
\begin{align}
    \hat U\ts{I}
    &=
        \mathcal{T}_t 
        \exp 
        \kako{
            -\ii \int_{\mathbb{R}} \dd t\,\hat H\ts{I}^t(t)
        } \,,
\end{align}
where $\mathcal{T}_t$ is a time-ordering symbol with respect to the common time $t$.

One can then use a perturbative analysis and obtain the final density matrix of a joint system $\mathcal{H}\ts{A} \otimes \mathcal{H}\ts{B}$, where $\mathcal{H}_j$ is a Hilbert space for detector-$j$. 
Assuming a small coupling strength, $\lambda \ll 1$,  the Dyson series expansion of $\hat U\ts{I}$ reads 
\begin{subequations}
\begin{align}
    \hat U\ts{I}
    &=
        \id + \hat U\ts{I}^{(1)} + \hat U\ts{I}^{(2)} + \mathcal{O}(\lambda^3)\,,\\
    \hat U\ts{I}^{(1)}
    &=
        -\ii \int_{-\infty}^\infty \dd t\,\hat H\ts{I}^t(t)\,,\\
    \hat U\ts{I}^{(2)}
    &=
        - \int_{-\infty}^\infty \dd t_1
        \int_{-\infty}^{t_1} \dd t_2\,
        \hat H\ts{I}^t(t_1) \hat H\ts{I}^t(t_2)\,.
\end{align}
\end{subequations}
By assuming that the initial state, $\rho_0$, of the detectors-field system is 
\begin{align}
    \rho_0
    &=
        \ket{g\ts{A}} \bra{g\ts{A}}
        \otimes 
        \ket{g\ts{B}} \bra{g\ts{B}}
        \otimes 
        \ket{0}\bra{0}\,,
\end{align}
where $\ket{0}$ is the vacuum state of the field, one finds the final total density matrix $\rho\ts{tot}$ after the interaction to be
\begin{align}
    \rho\ts{tot}
    &=
        \hat U\ts{I} \rho_0 \hat U\ts{I}^\dag \notag \\
    &=
        \rho_0 
        + 
        \rho^{(1,1)}
        +
        \rho^{(2,0)}
        +
        \rho^{(0,2)}
        +
        \mathcal{O}(\lambda^4)\,,
\end{align}
where $\rho^{(i,j)}=\hat U^{(i)} \rho_0 \hat U^{(j)\dagger}$ and all the odd-power terms of $\lambda$ vanish \cite{pozas2015harvesting}. 
Then the final density matrix of the detectors, $\rho\ts{AB}$, is obtained by tracing out the field part: $\rho\ts{AB}=\Tr_\phi[\rho\ts{tot}]$. 
By employing the basis $\ket{g\ts{A} g\ts{B}}=[1,0,0,0]^\top, \ket{g\ts{A} e\ts{B}}=[0,1,0,0]^\top, \ket{e\ts{A} g\ts{B}}= [0,0,1,0]^\top,  \ket{e\ts{A} e\ts{B}}=[0,0,0,1]^\top$, the density matrix $\rho\ts{AB}$ reads
\begin{align}
    \rho\ts{AB}
    &=
        \left[
        \begin{array}{cccc}
        1-\mathcal{L}\ts{AA}-\mathcal{L}\ts{BB} &0 &0 &\mathcal{M}^*  \\
        0 &\mathcal{L}\ts{BB} &\mathcal{L}\ts{AB}^* &0  \\
        0 &\mathcal{L}\ts{AB} &\mathcal{L}\ts{AA} &0  \\
        \mathcal{M} &0 &0 &0 
        \end{array}
        \right]
        + \mathcal{O}(\lambda^4)\,, \label{eq:density matrix}
\end{align}
where
\begin{subequations}
    \begin{align}
        \mathcal{L}_{ij}
        &=
            \lambda^2
            \int_{\mathbb{R}} \dd \tau_i
            \int_{\mathbb{R}} \dd \tau_j'\,
            \chi_i(\tau_i) \chi_j(\tau_j')
            e^{ -\ii \Omega (\tau_i - \tau_j') } \notag \\
            &\qquad\qquad\qquad \qquad\times 
            W\big( \sx_i(\tau_i), \sx_j(\tau_j') \big)\,, 
        \label{eq:transitionprob} \\
    \mathcal{M}
    &=
        -\lambda^2
        \int_{\mathbb{R}} \dd \tau\ts{A}
        \int_{\mathbb{R}} \dd \tau\ts{B}\,
        \chi\ts{A}(\tau\ts{A}) \chi\ts{B}(\tau\ts{B})
        e^{ -\ii \Omega (\tau\ts{A} + \tau\ts{B}) } \notag \\
        &\hspace{5mm}\times 
        \big[ 
            \Theta \big( t(\tau\ts{A}) - t(\tau\ts{B}) \big)
            W \big( \sx\ts{A}(\tau\ts{A}), \sx\ts{B}(\tau\ts{B}) \big) \notag \\
            &\hspace{1cm}
            +
            \Theta \big( t(\tau\ts{B}) - t(\tau\ts{A}) \big)
            W \big( \sx\ts{B}(\tau\ts{B}), \sx\ts{A}(\tau\ts{A}) \big)
        \big]\,,
        \label{eq:M}
    \end{align}
    \label{eq:elements in density matrix}
\end{subequations}
where $\Theta(t)$ is the Heaviside step function and $W(\sx, \sx')\coloneqq \bra{0} \hat\phi(\sx) \hat \phi(\sx') \ket{0}$ is the vacuum Wightman function. 
In $(3+1)$-dimensional Minkowski spacetime, the Wightman function reads
\begin{align}
    W(\sx, \sx')
    &=
        -\dfrac{1}{4\pi^2}
        \dfrac{1}{ (t-t'-\ii \epsilon)^2 - (\bx - \bx')^2 }\,,
\end{align}
where $\epsilon$ is the UV cutoff. 
The elements $\mathcal{L}_{jj},\,j\in \{\AAA, \BB\}$ are the so-called transition probabilities (or response functions), which describe the probability of a detector transitioning from the ground to excited states, $\ket{g_j}\to \ket{e_j}$. 
The off-diagonal elements $\mathcal{M}$ and $\mathcal{L}\ts{AB}$ are responsible for harvesting entanglement and quantum mutual information, respectively, as we shall see in the next subsection.

Throughout this paper, we use a Gaussian switching function
\begin{align}
    \chi_j(\tau_j)
    &=
        e^{ -\tau_j^2/2\sigma^2 }\,,\label{eq:Gaussian switch}
\end{align}
where $\sigma > 0$ is the characteristic Gaussian width, which has the units of time. 
We will use $\sigma$ to make all quantities unitless (such as $\Omega \sigma$).

\subsection{Correlation measure}\label{subsec:correlation measure}

Let us introduce two measures for correlation: concurrence $\mathcal{C}\ts{AB}$ and quantum mutual information $I\ts{AB}$.

Concurrence    is a measure of entanglement \cite{Hill.Wootters.concurrence, Wotters1998entanglementmeasure}. 
Let $\rho\ts{AB}$ be the density matrix of a two-qubit system. 
We first define a matrix $\tilde{\rho}\ts{AB}$ as 
\begin{align}
    \tilde{\rho}\ts{AB}
    &\coloneqq
        (\hat \sigma_y \otimes \hat \sigma_y) \rho\ts{AB}^* (\hat \sigma_y \otimes \hat \sigma_y)\,,
\end{align}
where $\hat \sigma_y$ is the Pauli-$y$ operator and $\rho\ts{AB}^*$ is the  complex conjugate of $\rho\ts{AB}$. 
Then by denoting $w_i \in \R$, $(i=1,2,3,4)$ as eigenvalues of a Hermitian operator $\sqrt{ \sqrt{\rho\ts{AB}} \tilde{\rho}\ts{AB} \sqrt{\rho\ts{AB}} }$, the concurrence is defined as follows. 
\begin{align}
    \mathcal{C}\ts{AB}
    &\coloneqq
        \max \{ 0,\, w_1- w_2 - w_3 - w_4 \}\,, \\
        &\hspace{15mm}(w_1 \geq w_2 \geq w_3 \geq w_4)\,. \notag 
\end{align}
The concurrence is zero if and only if the state $\rho\ts{AB}$ is separable. 
In the case of our density matrix \eqref{eq:density matrix}, the concurrence is known to be 
\begin{align}
    \mathcal{C}\ts{AB}
    &=
        2 \max \{ 0,\, |\mathcal{M}| - \sqrt{ \mathcal{L}\ts{AA} \mathcal{L}\ts{BB} } \}
        +
        \mathcal{O}(\lambda^4)\,. \label{eq:concurrence}
\end{align}

Quantum mutual information \cite{nielsen2000quantum}, on the other hand, quantifies the amount of total correlation,  both classical and quantum. 
Quantum mutual information $I\ts{AB}$ between two qubits A and B up to second order in $\lambda$ is \cite{pozas2015harvesting}
\begin{align}
    I\ts{AB}
    &=
        \mathcal{L}_+ \ln \mathcal{L}_+
        + 
        \mathcal{L}_- \ln \mathcal{L}_- \notag \\
        &\hspace{5mm}-
        \mathcal{L}\ts{AA} \ln \mathcal{L}\ts{AA}
        -
        \mathcal{L}\ts{BB} \ln \mathcal{L}\ts{BB}
        + \mathcal{O}(\lambda^4)
        \,,\label{eq:mutual info}
\end{align}
where 
\begin{align}
    \mathcal{L}_\pm
    &\coloneqq
        \dfrac{1}{2}
        \kako{
            \mathcal{L}\ts{AA}
            +
            \mathcal{L}\ts{BB}
            \pm 
            \sqrt{ (\mathcal{L}\ts{AA}-\mathcal{L}\ts{BB})^2 + 4 |\mathcal{L}\ts{AB}|^2 }
        }. \label{eq:Lpm}
\end{align}

Note that, while concurrence \eqref{eq:concurrence} vanishes when the ``noise term'' $\sqrt{\mathcal{L}\ts{AA} \mathcal{L}\ts{BB} }$ exceeds the nonlocal element $|\mathcal{M}|$, the mutual information becomes zero when $|\mathcal{L}\ts{AB}|=0$. 
In addition, if $\mathcal{C}\ts{AB}=0$ but the mutual information is nonvanishing, then the extracted correlation by the detectors is either classical correlation or nondistillable entanglement.

\begin{figure*}[t]
\centering
\includegraphics[width=\textwidth]{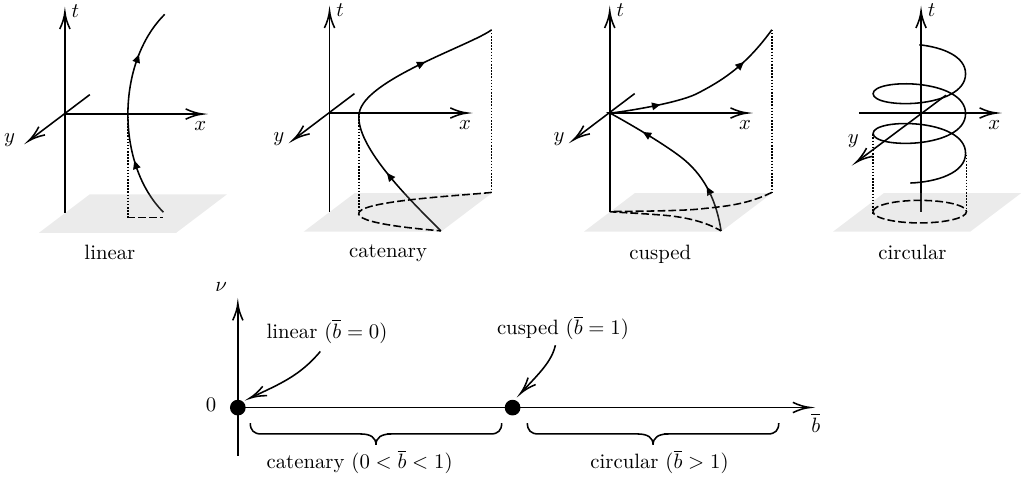}
\caption{Four trajectories characterized by $\bar b \equiv b/a$.}
\label{fig:four trajectories}
\end{figure*}

\section{Uniform acceleration trajectories}
\label{sec: uniform acceleration trajectories}

\subsection{Single detector trajectory classification}

The most well-known trajectory for a uniformly accelerating (i.e., $a=const.$) pointlike particle is   linear accelerated motion. 
However, Letaw pointed out that there are, in fact, five classes of uniformly accelerated trajectories, excluding the case where $a=0$. Along with the linear case, the other classes are circular, catenary, cusped, and helix \cite{Letaw.uniform.acceleration}. 
Consider a trajectory in $(3+1)$-dimensional Minkowski spacetime. 
Such a trajectory can be characterized by three geometric invariants: the curvature $a(\tau)$, which represents the magnitude of proper acceleration, the first torsion $b(\tau)$, and the second torsion (also known as hypertorsion) $\nu(\tau)$ of the worldline. 
The torsions $b(\tau)$ and $\nu(\tau)$ correspond to the proper angular velocities in a given tetrad frame \cite{Letaw.uniform.acceleration}. 
Assuming that these invariants are constants, the trajectory becomes stationary. 
In a nutshell, these motions are characterized by the following: 
\begin{enumerate}
    \item linear: $a\neq0, b=\nu=0$
    \item catenary: $a>b$, $\nu=0$
    \item cusped: $a=b, \nu=0$
    \item circular: $a<b, \nu=0$
    \item helix: $\nu \neq 0$
\end{enumerate}
In this subsection, we review these trajectories and consider the corresponding vacuum Wightman functions. 
We will suppress the UV cutoff $\epsilon$ for readability.

\subsubsection{Linear motion}
The linear acceleration motion of a detector is defined solely by the constant acceleration $a$, with all other parameters set to zero. 
The trajectory reads
\begin{align}
    \sx(\tau)
    &=
        \kako{
            \dfrac{1}{a}\sinh(a\tau), 
            \dfrac{1}{a}\cosh(a\tau),
            0,
            0
        }\,,
\end{align}
and the Wightman function along this trajectory is given by 
\begin{align}
    W\ts{lin}(\Delta \tau)
    &=
        -\dfrac{1}{4\pi^2}
        \dfrac{1}{ \dfrac{4}{a^2} \sinh^2 \kako{ \frac{a\Delta \tau}{ 2 } } }\,,
\end{align}
where $\Delta \tau \coloneqq \tau - \tau'$.

\subsubsection{Circular motion}
The circular trajectory is defined by $a$ and $b$ satisfying $a< b$. 
Let us begin with a commonly used trajectory 
\begin{align}
    \sx(\tau)
    &=
        \kako{
            \gamma \tau, 
            R\cos (\omega \gamma \tau),
            R\sin (\omega \gamma \tau),
            0
        }\,,
\end{align}
where $R, \omega$, and $\gamma$ are the radius of the circular motion, angular velocity, and the Lorentz factor defined as $\gamma \coloneqq 1/\sqrt{1-v^2}$. 
Here, $v\coloneqq R \omega (\leq 1)$ is the speed of the detector. 
Introducing the acceleration of the detector $a=R \omega^2 \gamma^2$, these parameters can be related by 
\begin{subequations}
    \begin{align}
    \omega
    &= 
        \sqrt{ \dfrac{a}{ (1+ a R ) R } }\,, \\
    \gamma
    &=
        \sqrt{ 1 + a R }\,, \\
    v 
    &= 
        \sqrt{ \dfrac{ a R }{ 1 + a R } }\,.
\end{align}
\end{subequations}
In terms of the acceleration $a$ and the torsion $b$, we can further express $\omega$ and $v$ as 
$$ 
\omega = b(1-a^2/b^2) \qquad  v=a/b
$$
respectively. 
The Wightman function is then 
\begin{align}
    W\ts{cir}(\Delta \tau) 
        = 
            -\dfrac{1}{4\pi^2}
            \dfrac{1}{\gamma^2 \Delta \tau^2 - 4 R^2\sin^2(\omega\gamma \Delta \tau/2)}\, .
\end{align}

\subsubsection{Cusped motion}\label{subsubsec:cusped}

Cusped motion is described by the acceleration and torsion with $a=b$. 
The trajectory reads 
\begin{align}
    \sx(\tau)
    &=
        \kako{
            \tau + \dfrac{1}{6} a^2 \tau^3,\,
            \dfrac{1}{2} a \tau^2,\,
            \dfrac{1}{6} a^2 \tau^3,\,
            0
        }\,,
\end{align}
and the corresponding Wightman function is 
\begin{align}
    W\ts{cus}(\Delta \tau)
    &=
        -\dfrac{1}{4\pi^2}
        \dfrac{1}{ \Delta \tau^2 + \dfrac{a^2}{12} \Delta \tau^4 }\,.\label{eq:cusped single Wightman}
\end{align}

\subsubsection{Catenary motion}\label{subsubsec:catenary}

Catenary motion can be characterized by $a$ and $b$ with $a>b$. The trajectory is given by 
\begin{align}
    \sx(\tau)
    &=
        \biggkako{
            \dfrac{a}{ a^2 - b^2 }
            \sinh \big(\sqrt{a^2 - b^2}\, \tau \big) , \notag \\
            &\quad
            \dfrac{a}{ a^2 - b^2 }
            \cosh \big( \sqrt{a^2 - b^2}\, \tau \big),
            \dfrac{b \tau}{\sqrt{a^2-b^2}},
            0
        }\,,
\end{align}
and the Wightman function reads
\begin{align}
    &W\ts{cat}(\Delta \tau) \notag \\
    &=
        -\dfrac{1}{4\pi^2}
        \dfrac{ 1 }
        {
            -\dfrac{b^2 \Delta \tau^2}{ a^2-b^2 }
            +
            \dfrac{4a^2}{ (a^2 - b^2)^2 }
            \sinh^2
            \big(
                \frac{ \sqrt{a^2-b^2} \Delta \tau }{2}
            \big)
        }.
\end{align}
We immediately see that  catenary motion reduces to the linear motion as $b\to 0$. 
Catenary motion also reduces to   cusped motion as $b\to a$ after a coordinate transformation consisting of a Lorentz boost a translation \cite{Good.Unruhlike.temperature2020}. 


\subsubsection{Helix motion}

Finally,   helix motion is a combination of circular and linear acceleration motions characterized by three parameters, $a, b$, and $\nu$: 
\begin{align}
    \sx(\tau)
    &=
        \biggkako{
            \dfrac{\mathcal{P}}{\Gamma_+} \sinh (\Gamma_+ \tau),
            \dfrac{ \mathcal{P} }{\Gamma_+} \cosh (\Gamma_+ \tau), \notag \\
            &\quad
            \dfrac{ \mathcal{Q} }{ \Gamma_- } \cos (\Gamma_- \tau),
            \dfrac{ \mathcal{Q} }{ \Gamma_- } \sin (\Gamma_- \tau)
        }\,,
\end{align}
where $\mathcal{P}\coloneqq \Xi/\Gamma$, $\mathcal{Q}\coloneqq ab/\Xi \Gamma$, and 
\begin{subequations}
    \begin{align}
    \Xi^2
    &\coloneqq
        \dfrac{1}{2} (\Gamma^2 + a^2 + b^2 + \nu^2)\,, \\
    \Gamma^2
    &\coloneqq
        \Gamma_+^2 + \Gamma_-^2\,, 
    \quad
    \Gamma_\pm^2 
    \coloneqq
        \sqrt{A^2 + B^2} \pm A\,, \\
    A
    &\coloneqq
        \dfrac{1}{2} (a^2 - b^2 - \nu^2)\,,
    \quad
    B
    \coloneqq
        a \nu\,.
\end{align}
\end{subequations}
The Wightman function reads
\begin{align}
    &W\ts{hel}(\Delta \tau)
    = \notag \\
        &-\dfrac{1}{4\pi^2}
        \dfrac{1}
        { 
            \dfrac{4 \mathcal{P}^2 }{ \Gamma_+^2 }
            \sinh^2 
            \kako{
                \frac{ \Gamma_+ \Delta \tau }{2}
            }
            - 
            \dfrac{4 \mathcal{Q}^2 }{ \Gamma_-^2 }
            \sin^2
            \kako{
                \frac{\Gamma_- \Delta \tau}{2}
            }
        }\,.
\end{align}
Note that the trajectory and the corresponding Wightman function reduce to the aforementioned trajectories when $\nu \to 0$. 
In this sense, the helix is the general motion that contains other motions.

\bigskip

\subsubsection{Wightman function at $\nu=0$}

We now turn our attention to the special case where $\nu=0$. 
Although the Wightman functions for linear, circular, catenary, and cusped motions may initially appear to take different forms, they can actually be expressed in a unified manner. 
Let $\bar b \equiv b/a$ with the condition that $a\neq 0$. 
The Wightman functions for all trajectories with $\nu=0$ can be written in the following compact form:

\begin{widetext}
    \begin{align}
    W_{\nu=0}(\Delta \tau)
    &=
        -\dfrac{1}{4\pi^2} 
        \dfrac{1}
        { 
            - \dfrac{\bar{b}^2}{1-\bar b^2} \Delta \tau^2
            + 
            \dfrac{4}{(1-\bar b^2)^2 a^2}
            \sinh^2
            \kako{
                \frac{ \sqrt{1-\bar b^2} a \Delta \tau }{2}
            }
        }\,.\label{eq:single Wightman general}
\end{align}
\end{widetext}
The parameter $\bar b$ serves to specify the particular trajectory, as illustrated in figure~\ref{fig:four trajectories}: linear ($\bar b=0$), catenary ($0<\bar b<1$), cusped ($\bar b=1$), and circular ($\bar b>1$). 
For  circular motion, we employ the identity $\sin(\ii x)=\ii \sinh(x)$. 
Note that one obtains the Wightman function for the cusped motion, as given in \eqref{eq:cusped single Wightman}, by performing a series expansion around $\bar b=1$.

The corresponding transition probability $\mathcal{L}_{jj}$, $j\in \{ \AAA, \BB \}$,  in \eqref{eq:density matrix} reads 
\begin{align}
    \mathcal{L}_{jj}
    &=
        \lambda^2 \sigma \sqrt{\pi} 
        \int_\R \dd u\,
        e^{ -u^2/4\sigma^2 }
        e^{-\ii \Omega u}
        W_{\nu=0}(u)\,. \label{eq:general acceleration transition}
\end{align}

\begin{figure*}[t]
\centering
\includegraphics[width=\textwidth]{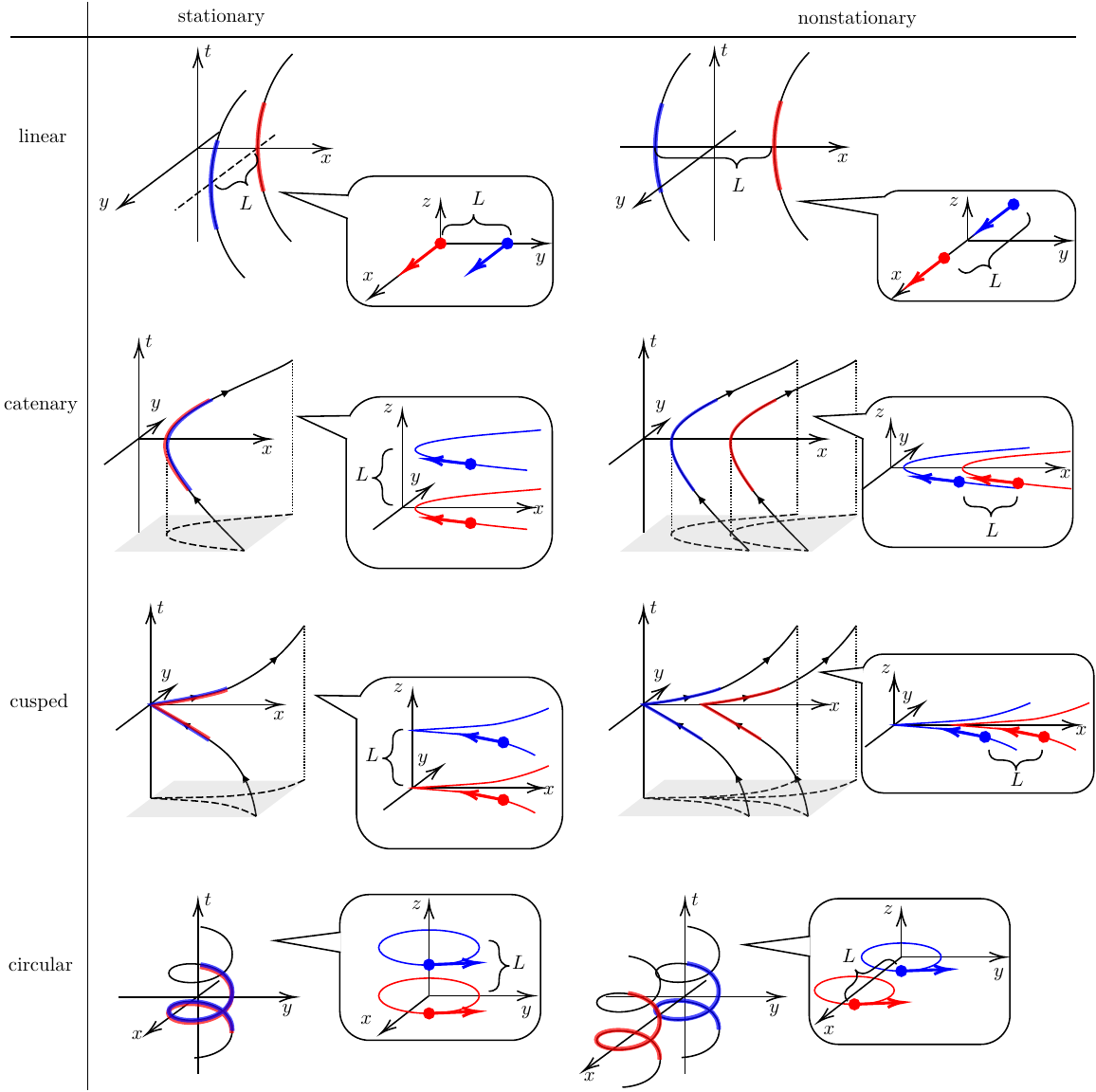}
\caption{Stationary and nonstationary configurations for four classes of uniformly accelerating detectors. 
Red and blue strips represent detectors A and B, respectively. }
\label{fig:two detectors scenarios}
\end{figure*}

\subsection{Two detectors in uniform acceleration}\label{subsec:two detector configs}
We now consider two UDW detectors A and B, both undergoing uniform acceleration motion. 
In particular, we categorize the detector configurations into two classes: stationary (time-translation invariant) and nonstationary scenarios.

\subsubsection{Stationary scenario}\label{subsubsec:time-trans invariant}

Consider two detectors undergoing the same uniform acceleration  (e.g., both  linearly accelerated). 
The Wightman function can be made time-translation invariant, meaning it depends only on the time difference $\Delta \tau \coloneqq \tau\ts{A} - \tau\ts{B}$, by imposing that the angle between the velocity vector of a detector and the spatial displacement vector from one detector to the other is time-independent. 
For example, two linearly accelerating detectors along the trajectories
\begin{subequations}
    \begin{align}
    \sx\ts{A}(\tau\ts{A})
    &=
        \kako{
            \dfrac{1}{a} \sinh(a\tau\ts{A}),
            \dfrac{1}{a} \cosh(a\tau\ts{A}),
            0,
            0
        }\,, \\
    \sx\ts{B}(\tau\ts{B})
    &=
        \kako{
            \dfrac{1}{a} \sinh(a\tau\ts{B}),
            \dfrac{1}{a} \cosh(a\tau\ts{B}),
            L,
            0
        }
\end{align}
\end{subequations}
give the following stationary Wightman function: 
\begin{align}
    &W\ts{lin}(\tau\ts{A}, \tau\ts{B}) 
    =
        -\dfrac{1}{4\pi^2}
        \dfrac{1}
        { 
            \dfrac{4}{a^2} \sinh^2
            \kako{
                \frac{a \Delta \tau}{2}
            }
            -L^2
        }\, 
\end{align}
where $L\coloneqq |\bx\ts{A} - \bx\ts{B}|$ is the spatial separation between the two detectors. 
As depicted in figure~\ref{fig:two detectors scenarios} top-left, each velocity vector of the detector is always perpendicular to the displacement vector $\bx\ts{AB}\coloneqq \bx\ts{A} - \bx\ts{B}$ throughout the interaction.

We can also construct stationary Wightman functions for other motions: \\

\textbf{circular}:
\begin{subequations}
    \begin{align}
        &\sx\ts{A}
        =
            \kako{
                \gamma \tau\ts{A}, 
                R\cos (\omega \gamma \tau\ts{A}),
                R\sin (\omega \gamma \tau\ts{A}),
                0
            }\,, \\
        &\sx\ts{B}
        =
            \kako{
                \gamma \tau\ts{B}, 
                R\cos (\omega \gamma \tau\ts{B}),
                R\sin (\omega \gamma \tau\ts{B}),
                L
            }\,.
    \end{align}
\end{subequations}
\textbf{cusped}:
\begin{subequations}
    \begin{align}
        &\sx\ts{A}
        =
            \kako{
                \tau\ts{A} + \dfrac{1}{6} a^2 \tau\ts{A}^3,\,
                \dfrac{1}{2} a \tau\ts{A}^2,\,
                \dfrac{1}{6} a^2 \tau\ts{A}^3,\,
                0
            }\,, \\
        &\sx\ts{B}
        =
            \kako{
                \tau\ts{B} + \dfrac{1}{6} a^2 \tau\ts{B}^3,\,
                \dfrac{1}{2} a \tau\ts{B}^2,\,
                \dfrac{1}{6} a^2 \tau\ts{B}^3,\,
                L
            }\,,
    \end{align}
\end{subequations}
\textbf{catenary}:
\begin{subequations}
    \begin{align}
        &\sx\ts{A}
        =
            \biggkako{
                \dfrac{a}{ a^2 - b^2 }
                \sinh \big( \sqrt{a^2 - b^2}\, \tau\ts{A} \big), \notag \\
                &\hspace{1cm}
                \dfrac{a}{ a^2 - b^2 }
                \cosh \big( \sqrt{a^2 - b^2}\, \tau\ts{A} \big),
                \dfrac{b \tau\ts{A}}{\sqrt{a^2-b^2}},
                0
            }\,, \\
        &\sx\ts{B}
        =
            \biggkako{
                \dfrac{a}{ a^2 - b^2 }
                \sinh \big( \sqrt{a^2 - b^2}\, \tau\ts{B} \big), \notag \\
                &\hspace{1cm}
                \dfrac{a}{ a^2 - b^2 }
                \cosh \big( \sqrt{a^2 - b^2}\, \tau\ts{B} \big),
                \dfrac{b \tau\ts{B}}{\sqrt{a^2-b^2}},
                L
            }\,,
    \end{align}
\end{subequations}
As for  a single detector, the Wightman functions along the trajectories given above take the following compact form: 
\begin{widetext}
    \begin{align}
    W\ts{s}(\tau\ts{A}, \tau\ts{B})
    \equiv 
        W\ts{s}(\Delta \tau)
    =
        -\dfrac{1}{4\pi^2} 
        \dfrac{1}
        { 
            - \dfrac{\bar b^2}{1-\bar b^2} \Delta \tau^2
            + 
            \dfrac{4}{(1-\bar b^2)^2 a^2}
            \sinh^2
            \kako{
                \frac{ \sqrt{1-\bar b^2} a \Delta \tau }{2}
            }
            -L^2
        }\,, \label{eq:stationary Wightman}
\end{align}
\end{widetext}
where $\Delta \tau \coloneqq \tau\ts{A} - \tau\ts{B}, \bar b \equiv b/a$, and the subscript `s' stands for stationary. 
Since the Wightman function depends only on $\Delta \tau$, the elements in the density matrix \eqref{eq:density matrix}, $\mathcal{M}$ and $\mathcal{L}\ts{AB}$, can be simplified to single integrals when the Gaussian switching function \eqref{eq:Gaussian switch} is used: 
\begin{subequations}
    \begin{align}
        \mathcal{M}
        &= 
            -2\lambda^2 \sigma \sqrt{\pi} e^{ -\Omega^2 \sigma^2 }
            \int_0^\infty \dd u\,
            e^{ -u^2/4\sigma^2 } W\ts{s}(u)\,, \\
        \mathcal{L}\ts{AB}
        &=
            \lambda^2 \sigma \sqrt{\pi} 
            \int_\R \dd u\,
            e^{ -u^2/4\sigma^2  }
            e^{-\ii \Omega u}
            W\ts{s}(u)\,.
    \end{align}\label{eq:elements for stationary}
\end{subequations}
Here, we used the fact that the Heaviside step function in \eqref{eq:M} can be written as $\Theta(t(\tau\ts{A}) - t(\tau\ts{B}))=\Theta (\tau\ts{A} - \tau\ts{B})$ for any of the uniform acceleration scenarios mentioned earlier. 

We note that all stationary configurations can only be realized in $(3+1)$ dimensions, with  the exception of the linear configuration.

\subsubsection{Nonstationary scenario}

One can also consider configurations similar to those in section \ref{subsubsec:time-trans invariant}, where the Wightman function depends not only on $\Delta \tau$ but also on $\Delta_+ \tau \coloneqq \tau\ts{A} + \tau\ts{B}$. 
In this case, the Wightman function is no longer time-translation invariant (hence, nonstationary).

In particular, consider two linearly accelerating UDW detectors whose trajectories are given by 
\begin{subequations}
    \begin{align}
    \sx\ts{A}(\tau\ts{A})
    &=
        \kako{
            \dfrac{1}{a} \sinh(a\tau\ts{A}),
            \dfrac{1}{a} \cosh(a\tau\ts{A}) + L,
            0,
            0
        }\,, \\
    \sx\ts{B}(\tau\ts{B})
    &=
        \kako{
            \dfrac{1}{a} \sinh(a\tau\ts{B}),
            \dfrac{1}{a} \cosh(a\tau\ts{B}),
            0,
            0
        }\,.
\end{align}
\end{subequations}
The correlation harvesting protocol along these trajectories was examined in \cite{Liu:2021dnl, ManarKenMutual}. 
The corresponding Wightman function reads 
\begin{widetext}
    \begin{align}
        W\ts{lin}(\tau\ts{A}, \tau\ts{B}) 
        &=
            -\dfrac{1}{4\pi^2}
            \dfrac{1}
            { 
                \dfrac{4}{a^2} \sinh^2
                \kako{
                    \frac{a \Delta \tau}{2}
                }
                -L^2
                -\dfrac{4L}{a} 
                \sinh 
                \kako{
                    \frac{a \Delta \tau}{2}
                }
                \sinh 
                \kako{
                    \frac{a \Delta_+ \tau}{2}
                }
            }\,.
    \end{align}
\end{widetext}
The term $\Delta_+\tau$ comes from the fact that the angle between the velocity vector and the displacement vector is time-dependent ($0^\circ$ or $180^\circ$).

Similarly, other uniformly accelerating trajectories that yield a nonstationary Wightman function are \\

\textbf{circular}:
\begin{subequations}
    \begin{align}
        &\sx\ts{A}
        =
            \kako{
                \gamma \tau\ts{A}, 
                R\cos (\omega \gamma \tau\ts{A}) + L,
                R\sin (\omega \gamma \tau\ts{A}),
                0
            }\,, \\
        &\sx\ts{B}
        =
            \kako{
                \gamma \tau\ts{B}, 
                R\cos (\omega \gamma \tau\ts{B}),
                R\sin (\omega \gamma \tau\ts{B}),
                0
            }\,.
    \end{align}
\end{subequations}
\textbf{cusped}:
\begin{subequations}
    \begin{align}
        &\sx\ts{A}
        =
            \kako{
                \tau\ts{A} + \dfrac{1}{6} a^2 \tau\ts{A}^3,\,
                \dfrac{1}{2} a \tau\ts{A}^2 + L,\,
                \dfrac{1}{6} a^2 \tau\ts{A}^3,\,
                0
            }\,, \\
        &\sx\ts{B}
        =
            \kako{
                \tau\ts{B} + \dfrac{1}{6} a^2 \tau\ts{B}^3,\,
                \dfrac{1}{2} a \tau\ts{B}^2,\,
                \dfrac{1}{6} a^2 \tau\ts{B}^3,\,
                0
            }\,,
    \end{align}
\end{subequations}
\textbf{catenary}:
\begin{subequations}
    \begin{align}
        &\sx\ts{A}
        =
            \biggkako{
                \dfrac{a}{ a^2 - b^2 }
                \sinh \big( \sqrt{a^2 - b^2}\, \tau\ts{A} \big), \notag \\
                &\hspace{1cm}
                \dfrac{a}{ a^2 - b^2 }
                \cosh \big( \sqrt{a^2 - b^2}\, \tau\ts{A} \big) + L,
                \dfrac{b \tau\ts{A}}{\sqrt{a^2-b^2}},
                0
            }\,, \\
        &\sx\ts{B}
        =
            \biggkako{
                \dfrac{a}{ a^2 - b^2 }
                \sinh \big( \sqrt{a^2 - b^2}\, \tau\ts{B} \big), \notag \\
                &\hspace{1cm}
                \dfrac{a}{ a^2 - b^2 }
                \cosh \big( \sqrt{a^2 - b^2}\, \tau\ts{B} \big),
                \dfrac{b \tau\ts{B}}{\sqrt{a^2-b^2}},
                0
            }\,,
    \end{align}
\end{subequations}
The Wightman function for these nonstationary motions can be compactly expressed as 
\begin{widetext}
    \begin{align}
    &W\ts{ns}(\tau\ts{A}, \tau\ts{B})= \notag \\
    &
        -\dfrac{1}{4\pi^2} 
        \dfrac{1}
        { 
            - \dfrac{\bar b^2}{1-\bar b^2} \Delta \tau^2
            + 
            \dfrac{4}{(1-\bar b^2)^2 a^2}
            \sinh^2
            \kako{
                \frac{ \sqrt{1-\bar b^2} a \Delta \tau }{2}
            }
            -L^2
            -\dfrac{4L}{(1-\bar b^2)a}
            \sinh
            \kako{
                \frac{ \sqrt{1-\bar b^2} a \Delta \tau }{2}
            }
            \sinh
            \kako{
                \frac{ \sqrt{1-\bar b^2} a \Delta_+ \tau }{2}
            }
        }\,,\label{eq:nonstationary Wightman}
\end{align}
\end{widetext}
and possesses an additional term in the denominator compared to the stationary Wightman function \eqref{eq:stationary Wightman}. 
Here, the subscript `ns' designates nonstationary. 
Due to this additional term, the correlations harvested by nonstationary detectors will exhibit behavior similar to those of stationary detectors. 

Note that the presence of $\Delta_+\tau$ prevents us from reducing the double integrals in \eqref{eq:elements in density matrix} into single integrals.  Furthermore,   all nonstationary configurations can   be realized in $(2+1)$ dimensions, except for the helix case, which we are not considering.

\begin{figure*}[t]
\centering
\includegraphics[width=\textwidth]{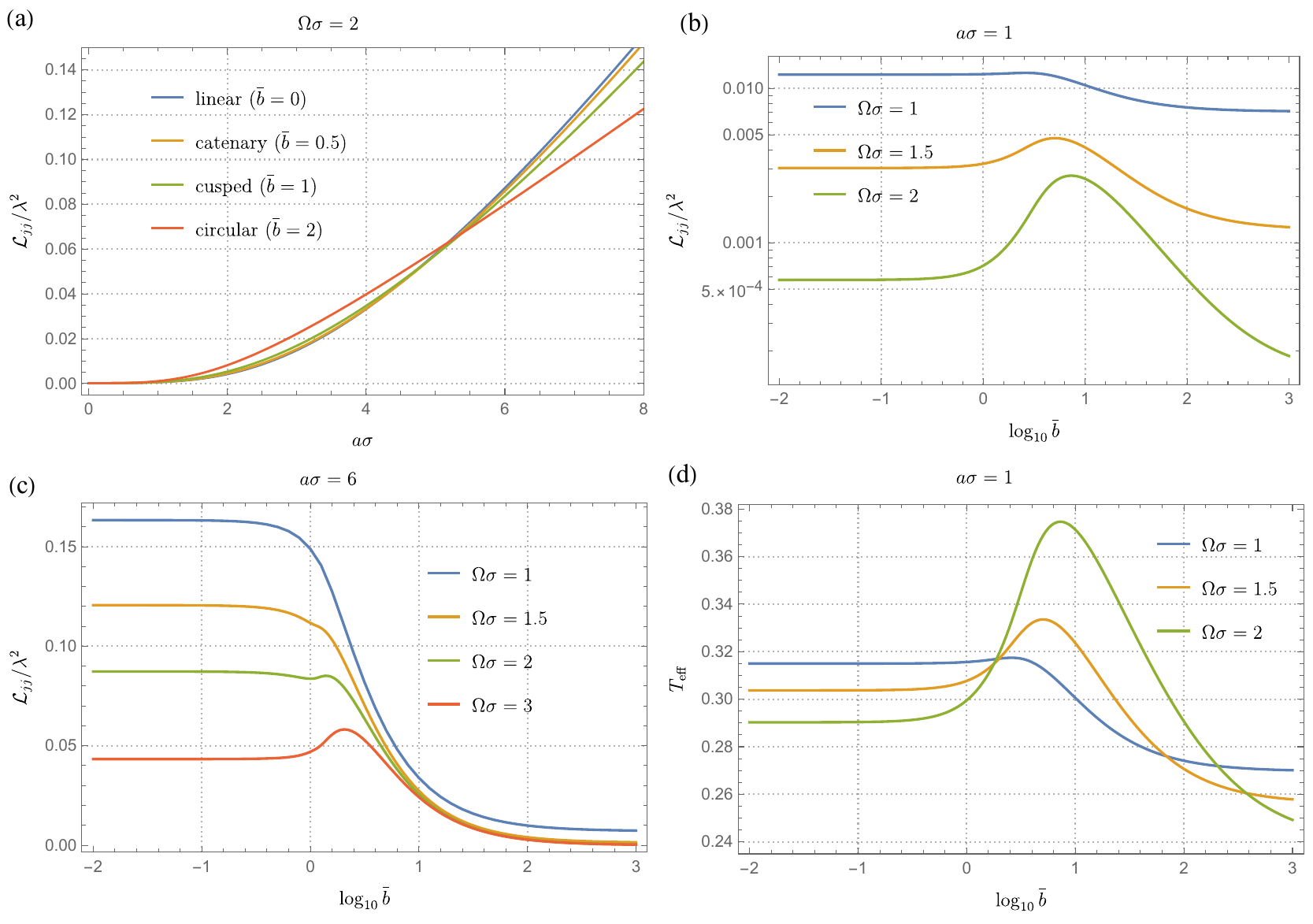}
\caption{
    (a) Transition probabilities $\mathcal{L}_{jj}/\lambda^2$ as a function of the magnitude of acceleration $a \sigma$ with $\Omega \sigma =2$. 
    (b) $\mathcal{L}_{jj}/\lambda^2$ as a function of $\log_{10}\bar b$ with $a\sigma=1$ and (c) with $a \sigma=6$. 
    (d) The effective temperature $T\ts{eff}$ as a function of $\log_{10}\bar b$ with $a \sigma =1$. 
}
\label{fig:transition prob}
\end{figure*}

\section{Numerical results}\label{sec:results}

Here, we numerically compute the concurrence \eqref{eq:concurrence} and quantum mutual information \eqref{eq:mutual info} harvested by two uniformly accelerating detectors by inserting the Wightman functions \eqref{eq:single Wightman general}, \eqref{eq:stationary Wightman} and \eqref{eq:nonstationary Wightman} into $\mathcal{L}_{ij}$ and $\mathcal{M}$ given in \eqref{eq:elements in density matrix}. 
For stationary detectors in \ref{subsubsec:time-trans invariant}, we utilize the expressions given by \eqref{eq:elements for stationary}.

\subsection{Transition probability of uniformly accelerating detectors}
\label{subsec:transition prob acceleration}

Let us begin by considering the transition probability $\mathcal{L}_{jj}$ for a uniformly accelerating detector. 
We are particularly interested in the cases of linear ($\bar b=0$), catenary ($0<\bar b<1$), cusped ($\bar b=1$), and circular ($\bar b>1$) motions, and their respective transition probabilities are given by \eqref{eq:general acceleration transition}. 
We consider $\mathcal{L}_{jj}/\lambda^2$ and write the parameters in units of $\sigma$, which makes the transition probability a function of three variables: $a \sigma, \bar b$, and $\Omega \sigma$. 
It is important to note that $\mathcal{L}\ts{AA}=\mathcal{L}\ts{BB}$, as we are assuming both detectors are identical.

Figure~\ref{fig:transition prob} depicts the transition probability $\mathcal{L}_{jj}/\lambda^2$ as a function of   the magnitude of acceleration $a \sigma$
for fixed $\Omega$ (panel (a))
and $\log_{10}\bar b$ for different values of the acceleration (panels (b), (c)).
In figure~\ref{fig:transition prob}(a), the transition probabilities for a detector with $\Omega \sigma=2$ in linear ($\bar b=0$), catenary ($\bar b=0.5$), cusped ($\bar b=1$), and circular ($\bar b=2$) motions are shown. 
We find that in all these cases, $\mathcal{L}_{jj}/\lambda^2$ increases with the acceleration $a\sigma$.\footnote{$\mathcal{L}_{jj}/\lambda^2$ is not guaranteed to \textit{monotonically} increase with $a\sigma$ for a finite interaction duration. For a detector in the linearly accelerated motion, such phenomenon is known as the (weak) anti-Unruh effect \cite{Brenna2016anti-unruh, Garay2016anti-unruh}.}

However, the relationship between the transition probabilities of detectors in different uniform motions is highly nontrivial. 
For instance, when a detector has $\Omega \sigma =2$ and $a\sigma \lesssim 5$, as depicted in figure~\ref{fig:transition prob}(a), a detector in circular motion with $\bar b=2$ shows the largest value of $\mathcal{L}_{jj}/\lambda^2$, whereas a detector in   linear motion ($\bar b=0$) shows the smallest. 
This relation, however, flips for $a \sigma \gtrsim 5$. 
We will numerically demonstrate that such relationships depend on the interplay between $a\sigma$ and $\Omega \sigma$.

\begin{figure*}[t]
\centering
\includegraphics[width=\textwidth]{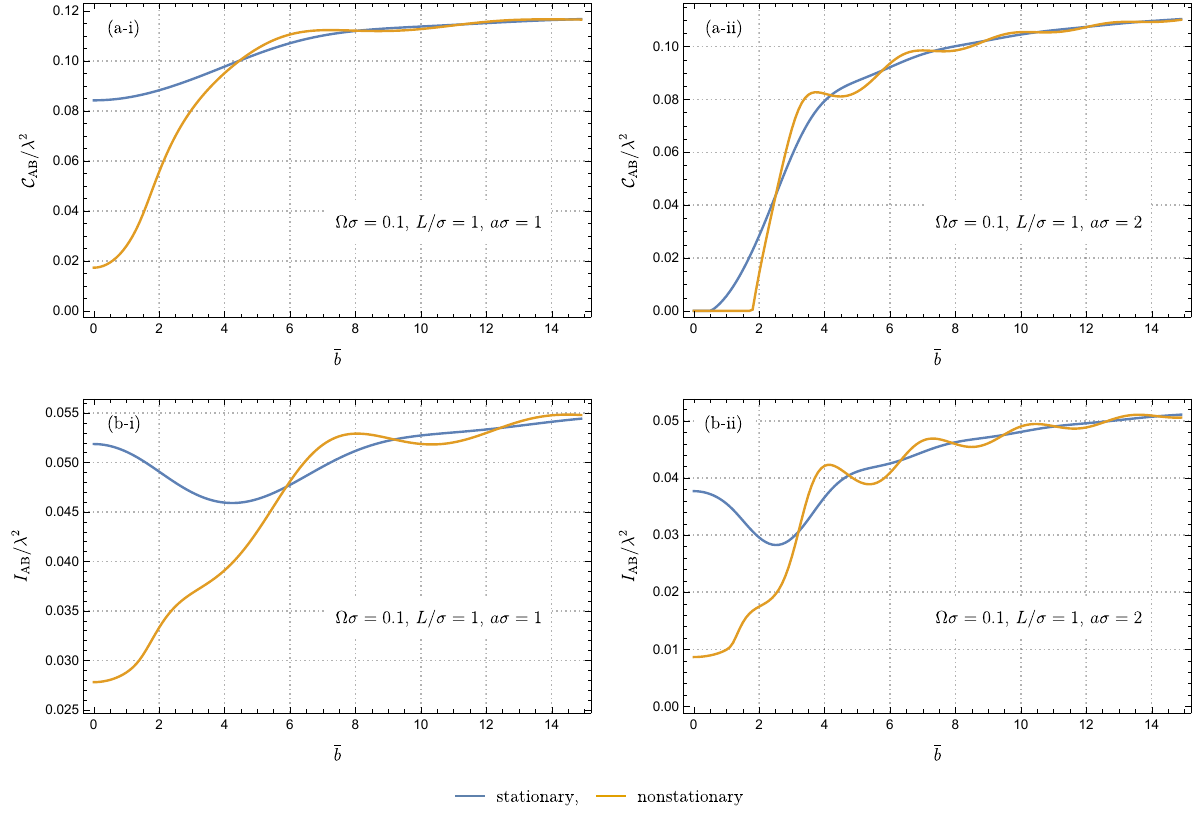}
\caption{
    Concurrence (a) and quantum mutual information (b) harvested by stationary and nonstationary detectors as a function of $\bar b$. 
    For each case, $\Omega \sigma =0.1$ and $L/\sigma=1$. 
    (a-i) and (b-i) are respectively $\mathcal{C}\ts{AB}/\lambda^2$ and $I\ts{AB}/\lambda^2$ when $a\sigma=1$, while (a-ii) and (b-ii) are respectively $\mathcal{C}\ts{AB}/\lambda^2$ and $I\ts{AB}/\lambda^2$ when $a\sigma=2$. 
}
\label{fig:FigConcMIvsbeta}
\end{figure*}

In figure~\ref{fig:transition prob}(b), the magnitude of the acceleration is fixed at $a\sigma =1$, and $\mathcal{L}_{jj}/\lambda^2$ is plotted as a function of $\log_{10}\bar b$. 
Each curve in this figure corresponds to a different value of $\Omega \sigma$, with the curve for $\Omega \sigma =2$ corresponding to figure~\ref{fig:transition prob}(a) at $a\sigma =1$. 
For each value of $\Omega \sigma$ in \ref{fig:transition prob}(b), the transition probability has a peak for $\log_{10}\bar b>0$ (i.e., $\bar b>1$), and then decreases with increasing $\log_{10}\bar b$, becoming smaller than the value for the linear case ($\log_{10}\bar b \to -\infty$). 
This means that $\mathcal{L}_{jj}/\lambda^2$ at $a \sigma = 1$ in figure~\ref{fig:transition prob}(a) increases with $\bar b$ until it reaches a maximum and then decreases. 
We note that the presence of the peak is contingent on larger values of $\Omega \sigma$ relative to $a \sigma$; 
In fact, the peak does not appear for smaller energy gaps, in which case the transition probability monotonically decreases with $\bar b$, as shown in figure~\ref{fig:transition prob}(b). 
This trend is further illustrated in figure~\ref{fig:transition prob}(c), where $a\sigma=6$ is chosen. 
In this scenario, the peak is nonexistent for $\Omega \sigma =1$ and 1.5 (as well as for $\Omega \sigma < 1$), but becomes manifest when $\Omega \sigma \gtrsim 2$. 
Thus we infer that detectors with smaller energy gaps $\Omega \sigma$ compared to $a \sigma$ do not have a peak in $\mathcal{L}_{jj}(\bar b)/\lambda^2$.

The behavior of $\mathcal{L}_{jj}/\lambda^2$ is related to the concept of the ``effective temperature'' perceived by a detector. 
For now, let us denote the transition probability as $\mathcal{L}_{jj}(\Omega,\sigma)$. 
The effective temperature, $T\ts{eff}$, is defined as 
\begin{align}
    T\ts{eff}^{-1}
    &\coloneqq
        \dfrac{1}{\Omega} 
        \ln 
        \dfrac{ \mathcal{L}_{jj}(-\Omega,\sigma)/\lambda^2 \sigma }{ \mathcal{L}_{jj}(\Omega,\sigma)/\lambda^2 \sigma }\,
\label{eq:effective temp}
\end{align}
where this formula is derived in the Appendix. We divide $\mathcal{L}_{jj}(\Omega,\sigma)$ by $\sigma$ so that it is well defined in the long interaction limit, $\sigma \to \infty$ \cite{Fewster.Waiting.Unruh, Garay2016anti-unruh}. 
Note that if the Wightman function obeys the   Kubo-Martin-Schwinger (KMS) condition \cite{Kubo1957thermality, Martin-Schwinger1959thermality}, then the effective temperature converges to the KMS temperature (which is the temperature of the field formally defined in quantum field theory) in the limit $\sigma \to \infty$. 
However, in the case of finite interaction duration, the effective temperature is  an estimator for the actual field temperature. 
For a detector in a uniform acceleration motion, the effective temperature for each scenario has been examined in, e.g., \cite{Bell.circular.1983, Bell.Leinaas.1987, Unruh.acceleration.rad.electron.1998, Juarez.Asymptotic.states.2019, Good.Unruhlike.temperature2020, CircularTemperaturesUnruh, bunney2023circular}.

We plot the effective temperature $T\ts{eff}$ as a function of $\log_{10}\bar b$ when $\sigma =1$ and $a \sigma =1$ in figure~\ref{fig:transition prob}(d), which corresponds to figure~\ref{fig:transition prob}(b). 
We see that the locations of the peaks in $T\ts{eff}$ align with those of $\mathcal{L}_{jj}(\Omega )$ in \ref{fig:transition prob}(b). 
This suggests that, for a given acceleration and energy gap, a detector in circular motion within a certain range of $\log_{10}\bar b$ can register higher effective temperatures than those in other types of motion. 
However, as $\bar b \to \infty$, which corresponds to the speed of a detector in circular motion with $v\ts{circ}(=\bar b^{-1}) \to 0$, the temperature becomes colder.

\subsection{Concurrence and quantum mutual information between uniformly accelerating detectors}
\label{subsec:Conc and MI vs a and b}

We now move on to the correlation harvesting protocol using two uniformly accelerating detectors, exploring both stationary and nonstationary configurations as described in section \ref{subsec:two detector configs}.

We first examine the difference between the stationary and nonstationary configurations by plotting concurrence $\mathcal{C}\ts{AB}/\lambda^2$ and quantum mutual information $I\ts{AB}/\lambda^2$ as a function of $\bar b$ in figure~\ref{fig:FigConcMIvsbeta}. 
In these plots, we fix $\Omega \sigma =0.1$ and $L/\sigma=1$, and consider $a\sigma=1$ and $a\sigma=2$. 
We notice two characteristics: 
(i) In the vicinity of $\bar b \approx 0$, stationary detectors consistently harvest greater correlations than  nonstationary detectors, for both concurrence and mutual information. 
(ii) As $\bar b$ becomes larger, both plots begin to oscillate with $\bar b$, and the curve representing correlations harvested by   nonstationary detectors oscillates around the curve for the stationary case. 
The frequency of the oscillation increases as $a\sigma$ grows.

These observations can be traced back to the form of the Wightman functions \eqref{eq:stationary Wightman} and \eqref{eq:nonstationary Wightman}. 
Let us recall that the denominators of these expressions contain $\sinh(x)$ when $\bar b \in [0,1)$ and transform into $\sin(x)$ when $\bar b>1$. 
Therefore, within the range $\bar b \in [0,1)$, the correlations are characterized by an exponential pattern, while for $\bar b >1$, an oscillatory behavior emerges. 
These traits explain the observation above. 
In particular, the suppression of correlations near $\bar b \approx 0$ for  nonstationary detectors can be attributed to an additional term in the denominator of  \eqref{eq:nonstationary Wightman}, which is absent in the stationary Wightman function \eqref{eq:stationary Wightman}. 
This extra term diminishes the amount of harvested correlations relative to the stationary scenario, and simultaneously gives rise to the oscillations noticed in the nonstationary case around the stationary one.

\begin{figure*}[t]
\centering
\includegraphics[width=\textwidth]{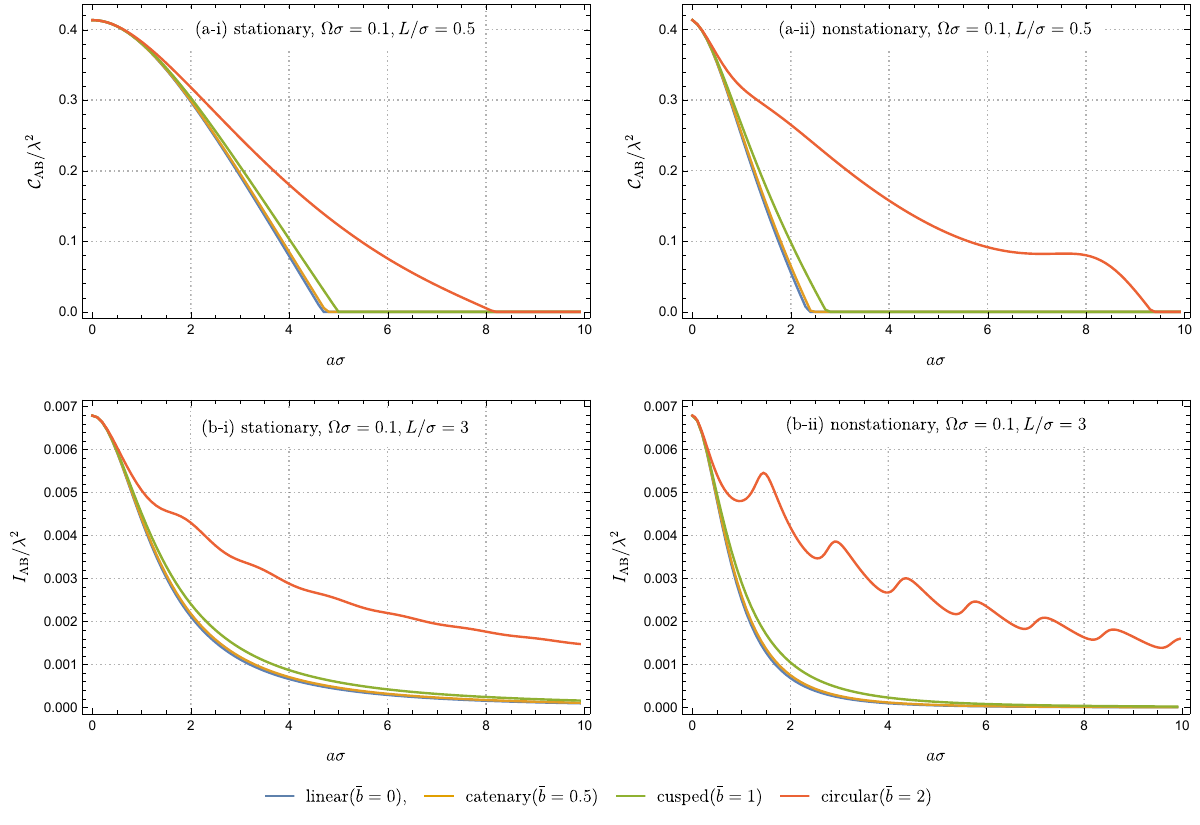}
\caption{Concurrence $\mathcal{C}\ts{AB}/\lambda^2$ with $\Omega \sigma=0.1, L/\sigma=0.5$ (a) and quantum mutual information $I\ts{AB}/\lambda^2$ with $\Omega \sigma=0.1, L/\sigma=3$ (b) as a function of the magnitude of acceleration $a\sigma$. 
(a-i) and (b-i) correspond to the stationary configuration while (a-ii) and (b-ii) show the nonstationary one, and each figure has four curves indicating four different motions. 
}
\label{fig:FigConcMIvsa}
\end{figure*}

We next examine the acceleration dependence of concurrence $\mathcal{C}\ts{AB}$ and quantum mutual information $I\ts{AB}$ as illustrated in figures~\ref{fig:FigConcMIvsa}(a) and (b), respectively. 
The stationary (figure~\ref{fig:FigConcMIvsa}(a-i) and (b-i)) and  nonstationary (figure~\ref{fig:FigConcMIvsa}(a-ii) and (b-ii)) configurations are depicted, and all four uniformly accelerated motions, linear ($\bar b=0$), catenary ($\bar b=0.5$), cusped ($\bar b=1$), and circular ($\bar b=2$) are shown in each figure.

As we pointed out earlier, the correlations harvested by nonstationary detectors for $\bar b \in [0,1)$ (figure~\ref{fig:FigConcMIvsa}(a-ii) and (b-ii)) decay with increasing $a\sigma$ faster than those extracted by the stationary detectors (figure~\ref{fig:FigConcMIvsa}(a-i) and (b-i)). 
Meanwhile, the correlations extracted by nonstationary detectors in circular motion ($\bar b>1$) (figure~\ref{fig:FigConcMIvsa}(a-ii) and (b-ii)) exhibit oscillatory behavior around the corresponding stationary curves (figure~\ref{fig:FigConcMIvsa}(a-i) and (b-i)).

Another observation we make is that, for both stationary and nonstationary configurations and for any value of $\bar b$, $\mathcal{C}\ts{AB}/\lambda^2$ becomes 0 at sufficiently high $a\sigma$. 
This can be attributed to the high transition probability at large $a\sigma$ as shown in figure~\ref{fig:transition prob}(a), leading to $|\mathcal{M}|<\sqrt{ \mathcal{L}\ts{AA} \mathcal{L}\ts{BB} }$ in \eqref{eq:concurrence}. 
Furthermore, the high accelerations prevent the detectors from extracting quantum mutual information, as $I\ts{AB}/\lambda^2 \to 0$ at $a\sigma \to \infty$ in figure~\ref{fig:FigConcMIvsa}(b). 
This indicates that any correlations cannot be harvested as $a\sigma \to \infty$ if the detectors are uniformly accelerated. 
These findings are consistent with previous results  \cite{Doukas.orbit.PhysRevA.81.062320, salton2015acceleration, Zhang.harvesting.circular, Liu.harvesting.reflecting.boundary, Liu:2021dnl, Liu.acceleration.vs.thermal, ManarKenMutual}, where linearly and circularly accelerated detectors are considered. 
Our paper extends these insights, providing a more general understanding that encompasses arbitrary uniformly accelerated motion.

\subsection{Genuine entanglement}
\label{subsec:genuine entanglement}

\begin{figure}[t]
\centering
\includegraphics[width=\linewidth]{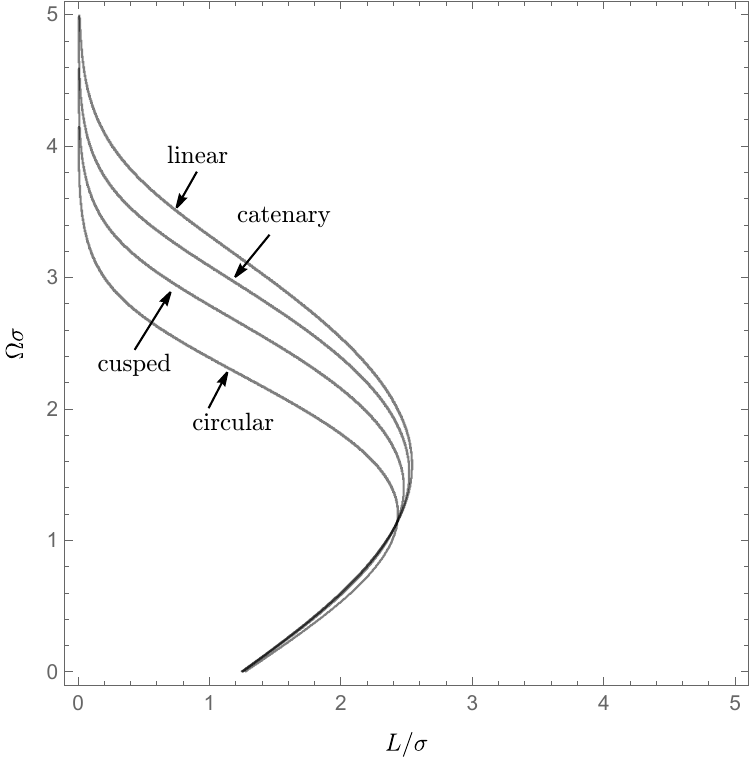}
\caption{The boundaries between $\mathcal{C}\ts{AB}>0$ and $\mathcal{C}\ts{AB}=0$ for the four stationary trajectories as a function of the proper separation $L/\sigma$ and the energy gap $\Omega \sigma$. 
Here, linear ($\bar b=0$), catenary ($\bar b=0.5$), cusped ($\bar b=1$), and circular ($\bar b=2$) with $a \sigma=1$ are depicted. 
Concurrence is nonzero in the left region of each curve. }
\label{fig:contour plot}
\end{figure}

We finally consider how much of entanglement is coming from the quantum field. 
It is known that the Wightman function can be decomposed into two parts: the anticommutator and the commutator of the field operator. 
The anticommutator part $\braket{ \{ \hat \phi(\sx), \hat \phi(\sx') \} }_{\rho_{\phi}}$ (also known as the Hadamard function), where $\braket{ \cdot }_{\rho_\phi}$ is the expectation value with respect to the field state $\rho_\phi$, depends on the state of the field $\rho_\phi$. 
Conversely, the commutator part $\braket{[ \hat \phi(\sx), \hat \phi(\sx') ]}_{\rho_\phi}=[ \hat \phi(\sx), \hat \phi(\sx') ] \in \mathbb{C}$ (also known as the Pauli-Jordan function) is state-independent. 
This means that even if the field state is not entangled, the commutator part in the Wightman function allows detectors to be entangled with each other. 
Such entanglement does not come from   preexisting entanglement in the field; rather it is associated with communication between the detectors, and thus we cannot say 
(for an unentangled field state) that
entanglement is `extracted' from the field if the commutator part is the only contribution \cite{TjoaSignal}. 
We say that entanglement is harvested if the anticommutator contribution in the element $\mathcal{M}$ is nonzero, and in particular we qualify the harvested entanglement as being \textit{genuine} if the commutator part in $\mathcal{M}$ is zero.   
Microcausality tells us that the two detectors can harvest genuine entanglement if they are causally disconnected. 
Here, we explore the circumstances under which  two uniformly accelerating detectors can extract genuine entanglement from the field. 
Remarkably we find that this can be possible even if the detectors are in causal contact.

We begin by plotting the concurrence $\mathcal{C}\ts{AB}/\lambda^2$ as a function of the proper separation $L/\sigma$ between the detectors and the energy gap $\Omega \sigma$ in figure~\ref{fig:contour plot}. 
The respective curves correspond to linear ($\bar b=0$), catenary ($\bar b=0.5$), cusped ($\bar b=1$), and circular motions ($\bar b=2$) in the stationary configurations depicted in figure~\ref{fig:two detectors scenarios}. 
The left region of each curve represents the parameters $(L/\sigma, \Omega \sigma)$ that enable the detectors to become entangled, manifest as   $\mathcal{C}\ts{AB}/\lambda^2>0$. 
Conversely, the right region corresponds to $\mathcal{C}\ts{AB}/\lambda^2=0$. 
Therefore, the stationary linear configuration ($\bar b=0$) has the broadest parameter space that leads to $\mathcal{C}\ts{AB}/\lambda^2>0$ compared to any other stationary configurations.

It has been  shown \cite{pozas2015harvesting} that two detectors at rest in Minkowski spacetime with a Gaussian switching function can be entangled with an arbitrary detector separation $L/\sigma$ if the energy gap $\Omega \sigma$ is large enough. 
However, we see that this is not the case for uniformly accelerating detectors -- they can be entangled only when they are close to each other, no matter how large $\Omega \sigma$ is.

We further ask how much  entanglement stems from the anticommutator and commutator parts in the Wightman function. 
To see this, let us decompose the Wightman function as 
\begin{align}
    W(\sx, \sx')
    &=
        \text{Re}[ W(\sx, \sx') ]
        + 
        \ii\, \text{Im}[ W(\sx, \sx') ] \,,
\end{align}
where 
\begin{subequations}
    \begin{align}
        2 \text{Re}[ W(\sx, \sx') ]
        &=
            \braket{0|\{ \hat \phi(\sx), \hat \phi(\sx') \}|0}\,, \\
        2 \text{Im}[ W(\sx, \sx') ]
        &=
            -\ii\, [ \hat \phi(\sx), \hat \phi(\sx') ]\,.
    \end{align}
\end{subequations}
Then the matrix element $\mathcal{M}$ can be decomposed into 
\begin{align}
    \mathcal{M}
    &=
        \mathcal{M}_+ + \ii \mathcal{M}_-\,,
\end{align}
where $\mathcal{M}_+$ and $\mathcal{M}_-$ are \eqref{eq:M} with the Wightman function being replaced by $\text{Re}[ W(\sx, \sx') ]$ and $\text{Im}[ W(\sx, \sx') ]$, respectively. 
$\mathcal{M}_+$ contains the information about the genuine entanglement whereas $\mathcal{M}_-$ is state-independent and does not necessarily exhibit the preexisted entanglement in the field. 
For the stationary detectors, these expressions can be simplified to single integral forms: 
\begin{subequations}
    \begin{align}
    \mathcal{M}_+
    &=
        -\lambda^2 \sigma \sqrt{\pi} e^{ -\Omega^2 \sigma^2 }
        \int_0^\infty \dd u\,
        e^{ -u^2/4\sigma^2 } W\ts{s}(u) \notag \\
        &\quad
        -\lambda^2 \sigma \sqrt{\pi} e^{ -\Omega^2 \sigma^2 }
        \int_0^\infty \dd u\,
        e^{ -u^2/4\sigma^2 } W\ts{s}^*(u)\,, \\
    \mathcal{M}_-
    &=
        \ii \lambda^2 \sigma \sqrt{\pi} e^{ -\Omega^2 \sigma^2 }
        \int_0^\infty \dd u\,
        e^{ -u^2/4\sigma^2 } W\ts{s}(u) \notag \\
        &\quad
        -\ii \lambda^2 \sigma \sqrt{\pi} e^{ -\Omega^2 \sigma^2 }
        \int_0^\infty \dd u\,
        e^{ -u^2/4\sigma^2 } W\ts{s}^*(u)\,.
\end{align}
\end{subequations}
We then define harvested concurrence $\mathcal{C}\ts{AB}^+$ and communication-assisted concurrence $\mathcal{C}\ts{AB}^-$ as \cite{TjoaSignal}
\begin{align}
    \mathcal{C}\ts{AB}^\pm 
    &\coloneqq
        2 \max \{ 0,\,|\mathcal{M}_\pm| - \sqrt{ \mathcal{L}\ts{AA} \mathcal{L}\ts{BB} } \}
        + \mathcal{O}(\lambda^4)\,.
\end{align}

\begin{figure*}[t]
\centering
\includegraphics[width=\linewidth]{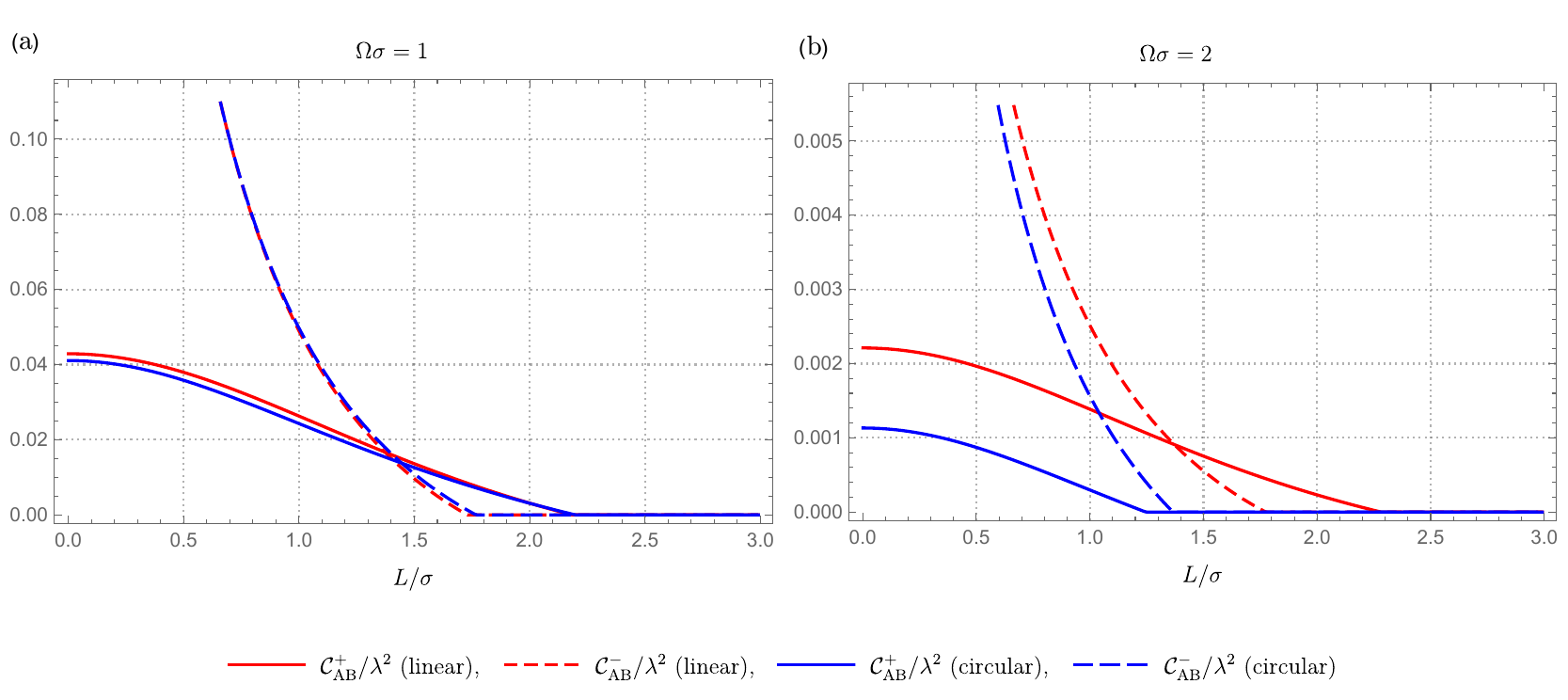}
\caption{Harvested and communication-assisted concurrence $\mathcal{C}\ts{AB}^\pm/\lambda^2$ for the stationary linear and circular configurations as a function of the detector separation $L/\sigma$. 
(a) When $\Omega \sigma=1$. 
The linear and circular cases are very similar and the detectors can harvest genuine entanglement near $L/\sigma=2$. 
(b) When $\Omega \sigma=2$. 
Although the linear case does not change much compared to the $\Omega \sigma=1$ case, the circular case cannot harvest genuine entanglement anymore. 
It is the mixture of the anticommutator and commutator contributions, or in the worst case $\mathcal{C}\ts{AB}^+/\lambda^2=0$ around $L/\sigma=1.3$. }
\label{fig:communication plot}
\end{figure*}

We plot $\mathcal{C}\ts{AB}^\pm/\lambda^2$ as a function of $L/\sigma$ in figure~\ref{fig:communication plot}. 
Here, we specifically choose the stationary linear ($\bar b=0$) and circular ($\bar b=2$) cases as a demonstration. 
We find that for $\Omega \sigma=1$ (figure~\ref{fig:communication plot}(a)), the detectors can harvest entanglement since $\mathcal{C}\ts{AB}^+/\lambda^2 > 0$. 
Most strikingly,  it is possible to extract genuine entanglement 
for $L/\sigma \in (1.5,2.2)$
since
$\mathcal{C}\ts{AB}^+ > 0$ while $\mathcal{C}\ts{AB}^- = 0$ in this region.
However, this is not always true as one can see from figure~\ref{fig:communication plot}(b) when $\Omega \sigma =2$. 
Here,   detectors in circular motion can encounter the case where $\mathcal{C}\ts{AB}^+=0$ while $\mathcal{C}\ts{AB}^- > 0$, which indicates that the generated entanglement after the interaction is purely coming from the communication and not from the field. 

However genuine entanglement can still be extracted in the linear configuration.

\section{Conclusion}
\label{sec: conclusion}

We carried out the correlation harvesting protocol using two uniformly accelerating Unruh-DeWitt (UDW) detectors in $(3+1)$-dimensional Minkowski spacetime. 
According to Letaw \cite{Letaw.uniform.acceleration},  trajectories with constant (nonzero) acceleration can be characterized by the magnitude of the acceleration and two torsion parameters, resulting in five classes: linear, catenary, cusped, circular, and helix motions. 
The first four of these classes of motion are confined to a two-dimensional spatial surface and can be regarded as specific cases of the helix motion. 
Since two-dimensional configurations are more amenable to experimental setups,
  we employed these four simpler motions for our analysis.

We first examined the transition probability of a single detector following the four trajectories in section~\ref{subsec:transition prob acceleration}. 
Utilizing a unified expression for the Wightman functions along these trajectories, we were able to explore the general characteristics that are common to all these motions. 
We found that the transition probabilities of these motions monotonically increase with the magnitude of acceleration. 
Moreover, we also evaluated the effective temperature---an estimator for the temperature as observed by a detector.

We then introduced another UDW detector to consider the correlation harvesting protocol in Sec.~\ref{subsec:Conc and MI vs a and b}. 
Two configurations were explored: stationary and nonstationary configurations. 
In the stationary configuration, detectors are separated in the direction perpendicular to their two-dimensional spatial planes of motion. 
Specifically, the displacement vector pointing from one detector to the other remains orthogonal to the velocity vectors of the detectors. 
In such a case, the Wightman function along the stationary configuration is time-translation invariant. 
On the other hand, in the nonstationary configuration, the displacement vector aligns parallel to the planes of motion. 
This makes the Wightman function nonstationary (i.e., not time-translation invariant). 
Moreover, while this Wightman function shares a common term with the stationary configuration, an additional term appears that specifically characterizes the nonstationary nature of this configuration.

We found that the harvested correlations---entanglement and total correlations---behave in a distinct manner depending on the motion of the detectors. 
Specifically, detectors in linear, catenary, and cusped motions within the nonstationary configuration gain fewer correlations compared to those in the stationary configuration. 
On the other hand, in the circular motion case, both configurations exhibit similar behavior. 
This difference can be attributed to the Wightman functions. 
For linear, catenary, and cusped motions, the Wightman function contain hyperbolic functions, leading to an exponential alteration of the results. 
In contrast, the  Wightman function for circular motion is governed by trigonometric functions.

We also looked into the acceleration dependence of the harvested correlations and concluded (not surprisingly) 
that high accelerations prevent the detectors from acquiring correlations from the field. 
This point is consistent with   previous papers \cite{Doukas.orbit.PhysRevA.81.062320, salton2015acceleration, Zhang.harvesting.circular, Liu.harvesting.reflecting.boundary, Liu:2021dnl, Liu.acceleration.vs.thermal, ManarKenMutual}, in which linearly and circularly accelerated detectors are considered. 
Our paper generalized these results to any uniformly accelerating detectors on two-dimensional spatial surfaces.

Finally, we focused on entanglement harvested by the detectors in the stationary configuration and   asked how much of entanglement is coming from the correlations preexisted in the field. 
To be precise, the entanglement coming from the commutator part of the Wightman function is state-independent, which suggests that the detectors can still be correlated even if the field is not entangled \cite{TjoaSignal}. 
Thus, it is important to examine how the anticommutator part of the Wightman function (which is state-dependent) contributes to the extracted correlations. 
One way to eliminate the commutator contribution is to use causally disconnected detectors. 
However, we found that the existence of acceleration prohibits us to extract correlations with detectors separated far away, no matter what the energy gap is. 
However we also found the striking result that 
detectors  in causal contact can harvest genuine entanglement in certain parameter regimes. 

Our results have important implications for experiment.  Attempts to realize the Unruh effect and correlation harvesting generally rely on using laser pulses to probe what are effectively two dimensional surfaces. 
To probe the effects of non-inertial motion on 
mutual information and entanglement
will therefore involve two detectors (two pulses) in nonstationary configurations, since only these can be realized in a two-dimensional setting. Experimental verification of the harvesting of genuine entanglement would be an exciting confirmation of our understanding of relativistic quantum information.

\section*{Acknowledgments}
This work was supported in part by the Natural Sciences and Engineering Research Council of Canada. 
KGY is thankful to Dr. Jorma Louko for elaborating on the relationship among the uniformly accelerating trajectories.

\appendix
\section{Effective temperature}
Here, we review the concept of effective temperature $T\ts{eff}$ and clarify its relation to the KMS temperature.

Let us first review the KMS temperature. 
In quantum theory with separable Hilbert spaces, a trace of an operator, $\Tr[\cdot]$ is well defined. 
This enables us to consider the Gibbs state at the inverse temperature $\beta$, $\rho=e^{-\beta \hat H}/Z$, where $Z\coloneqq \Tr[ e^{-\beta \hat H} ]$ is the partition function. 
This is what we consider a thermal state of a system.

However in QFT, a trace is generally not well defined. 
Instead, we identify the Kubo-Martin-Schwinger (KMS) state \cite{Kubo1957thermality, Martin-Schwinger1959thermality} as a thermal state in QFT. 
Specifically, if the field is in the KMS thermal state with respect to time $\tau$ at the inverse KMS temperature $\beta\ts{KMS}$, the Wightman function satisfies 
\begin{align}
    W(\Delta \tau - \ii \beta\ts{KMS})
    &=
        W(-\Delta \tau)\,,
\end{align}
where $\Delta \tau \coloneqq \tau - \tau'$. 
The Fourier transform of this equality with respect to $\Delta \tau$ reads 
\begin{align}
    \tilde{W} (-\omega) 
    &=
        e^{ \beta\ts{KMS} \omega } \tilde{W} (\omega)\,.
\end{align}
This equality in the Fourier domain is known as the detailed balance condition. 
Thus, the thermality of a quantum field is imprinted in these equalities.

The thermality can also be implemented in the transition probability of a UDW detector. 
Recall that the transition probability is written as 
\begin{align}
    \mathcal{L}
    &=
        \lambda^2
        \int_{\mathbb{R}} \dd \tau
        \int_{\mathbb{R}} \dd \tau'\,
        \chi(\tau) \chi(\tau')
        e^{ -\ii \Omega (\tau - \tau') } 
        W\big( \sx(\tau), \sx(\tau') \big)\,,
\end{align}
where the subscript in $\mathcal{L}_{jj},\,j\in \{ \AAA, \BB \}$ is omitted for simplicity. 
Let us assume that the switching function, $\chi(\tau)$, has a characteristic time length $\sigma$. 
In our paper, this is the Gaussian width in $\chi(\tau)=e^{ -\tau^2/2\sigma^2 }$. 
It is convenient to introduce a quantity related to the transition probability known as the \textit{response function} (divided by the characteristic time length) $\mathcal{F}(\Omega, \sigma)$: 
\begin{align}
    &\mathcal{L}
    =
        \lambda^2 \sigma \mathcal{F}(\Omega, \sigma)\,, \notag \\
    &\mathcal{F}(\Omega, \sigma) 
    \coloneqq \notag \\
        &\hspace{0.5cm}
        \dfrac{1}{\sigma }
        \int_{\mathbb{R}} \dd \tau
        \int_{\mathbb{R}} \dd \tau'\,
        \chi(\tau) \chi(\tau')
        e^{ -\ii \Omega (\tau - \tau') } 
        W\big( \sx(\tau), \sx(\tau') \big)\,.
\end{align}
If the field is in the KMS state and the switching function is a rapidly decreasing function such as a Gaussian function, then the response function in the long interaction limit obeys the detailed balance relation \cite{Fewster.Waiting.Unruh}: 
\begin{align}
    \lim_{\sigma \to \infty } \dfrac{ \mathcal{F}(-\Omega, \sigma) }{ \mathcal{F}(\Omega, \sigma) }
    = 
        e^{ \beta\ts{KMS} \Omega  }\,.
\end{align}
Note that this relation holds when the long interaction limit is taken. 
On the other hand, if $\sigma$ is not sufficiently long, the ratio of the response function (sometimes known as the excited-to-deexcited ratio) does not satisfy the detailed balance condition.

From this relation, one can define the \textit{effective temperature} as 
\begin{align}
    T\ts{eff}^{-1}
    \coloneqq 
        \dfrac{1}{\Omega }
        \ln 
        \dfrac{ \mathcal{F}(-\Omega, \sigma) }{ \mathcal{F}(\Omega, \sigma) }\,. \label{eq:effective temp resp}
\end{align}
Note that the effective temperature is not necessarily the KMS temperature. 
If the field is in the KMS state and the long interaction limit is taken, then the effective temperature becomes the KMS temperature. 
In this sense, the effective temperature is an estimator for the field's temperature.





\bibliography{ref}

\end{document}